\documentclass[aip,apl,amsmath,amssymb,preprint]{revtex4-1}

\usepackage{hyperref}

\usepackage{graphicx}
\usepackage{dcolumn}
\usepackage{bm}

\usepackage[utf8]{inputenc}
\usepackage[T1]{fontenc}
\usepackage{mathptmx}
\usepackage{etoolbox}
\usepackage{xcolor}

\usepackage[normalem]{ulem}

\makeatletter
\def\@email#1#2{%
 \endgroup
 \patchcmd{\titleblock@produce}
  {\frontmatter@RRAPformat}
  {\frontmatter@RRAPformat{\produce@RRAP{*#1\href{mailto:#2}{#2}}}\frontmatter@RRAPformat}
  {}{}
}%
\makeatother
\begin{document}
\raggedbottom

\newcommand{\gv}[1]{\textcolor{red}{GV: #1}}
\definecolor{ibg_c}{RGB}{0, 0, 255}   
\newcommand{\ibg}[1]{\textcolor{ibg_c}{#1}}


\title[]{Energy level alignment of vacancy-ordered halide double perovskites}

\author{I. B. Garba}
\author{G. Volonakis}
\email{yorgos.volonakis@univ-rennes.fr}

\affiliation{Univ Rennes, ENSCR, CNRS, ISCR (Institut des Sciences Chimiques de Rennes) UMR 6226, France}

\date{\today}

\begin{abstract}
Vacancy-ordered double perovskites have emerged as lead-free alternatives, offering remarkable stability and compositional tunability for optoelectronic applications. In this study, we provide first-principles insights into their electronic properties, surface stability, and energy level alignment using a non-empirical, dielectric-dependent hybrid functional. For a representative family of Cs$_2$MX$_6$ compounds, with M = Zr, Sn, Te, and X= Cl, Br, I, our calculations reveal that the predicted bulk electronic band gaps are in excellent agreement with those obtained using the state-of-the-art GW method, validating the accuracy of our approach. We investigate the stability of these materials under simulated experimental conditions, considering both the rich and poor chemical potentials of their precursor salts. Our results indicate distinct regions of surface energy stability that favor CsX terminations. In contrast, MX$_4$ terminations show in-gap surface states, which can act as trap states and reduce carrier lifetime. Finally, based solely on the intrinsic absolute energy levels, we identify promising candidates as charge transport and injection layers for typical photovoltaic and light-emitting applications. This study provides a detailed map of energy level alignment at Cs$_2$MX$_6$ surfaces, offering valuable design principles for the development of next-generation Cs$_2$MX$_6$-based optoelectronic devices.
\end{abstract}

\maketitle

Halide perovskite materials have attracted considerable attention, particularly for their use in solar cells with power conversion efficiencies (PCEs) exceeding 27\%~\cite{NREL_2025}, but also in light-emitting diodes (LEDs) achieving external quantum efficiencies (EQEs) over 20\%~\cite{Shi_2025}, and in sensitive photodetectors with >10$^{13}$ Jones~\cite{Sakhatskyi_2023, Zheng_2025}. Today, the best-performing halide perovskites are Pb-based, and concerns regarding lead toxicity and long-term stability have prompted the investigation of alternative compositions and structural variants for optoelectronic applications. An alternative family of materials is the so-called double perovskites, which crystallize in a similar lattice and are derived via heterovalent substitution of the divalent M cation in the AMX$_3$ structure, resulting in  A$_2$MM$'$X$_6$, where M and M$'$ are monovalent and trivalent metal cations, respectively~\cite{Slavney_2016, Volonakis_2017}. Within the same lattice, tetravalent substitution at the M site leads to the formation of vacancy-ordered double perovskites (VODPs) with formula A$_2$M$^{4+}$X$_6$, a subfamily of double perovskites featuring isolated MX$_6$ octahedra~\cite{Lee_2014, Maughan_2019}. The renewed interest in VODPs is primarily due to their environmental stability, low toxicity, and tunable optoelectronic properties, making them promising candidates for various applications~\cite{Lee_2014, Maughan_2019, Cucco_2021, Cucco_2024}. The first application of VODPs in solar cells was reported by Lee et al.\ in 2014~\cite{Lee_2014}, who introduced Cs$_2$SnI$_6$ as a cost-effective hole-transport layer (HTL) in dye-sensitized solar cells. Subsequent reports on mixed halides Cs$_2$SnI$_2$Br$_3$ showed low hole-transport resistance, a tunable band gap, and an improved conversion efficiency of 7\%~\cite{Kaltzoglou_2015, Kaltzoglou_2016}. While most VODPs are more stable than their AMX$_3$ counterparts, their typical wide band gaps and especially their large exciton binding energies~\cite{Cucco_2023,Kavanagh_2022} limit their suitability as photoactive solar materials. Their reported device efficiencies remain modest, with Cs$_2$SnI$_4$Br$_2$ at 2.1\%~\cite{Lee_2017} and Cs$_2$TiBr$_6$ at 3.3\%~\cite{Chen_2018a}. However, Schwartz et al.\ reported a remarkably high efficiency of 13.8\% using Cs$_2$PtI$_6$~\cite{Schwartz_2020}, yet not reproduced to-date. In contrast, VODPs have demonstrated high performance in LEDs~\cite{Tang_2025}, achieving a photoluminescence quantum yield of 95\% in Cs$_2$Sn$_{1-x}$Te$_{x}$Cl$_6$~\cite{Tan_2020}.

For optoelectronic applications, particularly solar cells and LEDs, the stability of the constituent layers and their energy-level alignment are critical factors that directly affect device efficiency~\cite{Schulz_2019, Zang_2024}. However, identifying optimal electron transport layers (ETLs) and hole transport layers (HTLs) that satisfy both electronic and stability requirements remains a significant challenge, because these requirements strongly depend on the choice of photoactive material. For example, in a perovskite-based solar cell, an ideal ETL should exhibit favorable energy alignment with the perovskite to facilitate electron transport while simultaneously blocking holes. Similarly, an optimal HTL must efficiently transport holes while preventing electron transport~\cite{Zang_2024}. Commonly used ETLs, such as TiO$_2$, SnO$_2$, and fullerene derivatives, are employed for their excellent transport properties, but energy-level mismatch with photoactive materials often requires interfacial modification to avoid charge-accumulation barriers that hinder charge transport~\cite{Luo_2020, Shi_2025}. Most HTLs are organic and hygroscopic; moreover, spiro-MeOTAD, the material of choice in high-performance devices, is particularly costly, comparable to platinum and gold in its high-purity form~\cite{Shahinuzzaman_2022}, posing both economic and stability challenges. To address these limitations, inorganic HTLs, such as CuI, Cu$_2$O, CuAlO$_2$, CuSCN, and NiO$_x$, have emerged as promising alternatives~\cite{Wang_2021, Shahinuzzaman_2022, Apergi_2020}. These materials offer low cost, high optical transmittance, and enhanced chemical stability under moisture and thermal stress~\cite{Zhu_2023, Wang_2021, Sun_2024}. Despite these advancements, large-scale commercialization of PSCs with inorganic HTLs still faces challenges, particularly in achieving optimal energy-level alignment between the inorganic HTL and the perovskite absorber. In this context, VODPs present an attractive alternative due to their ease of synthesis, environmental stability, and tunable band gaps~\cite{Maughan_2019}. Their transport properties and energy levels can be further optimized through halide mixing and M-site alloying~\cite{Zhang_2025, Phan_2025}, offering a versatile platform for engineering their optoelectronic properties to meet the specific requirements of targeted applications. Previous computational studies on the energy levels of Cs$_2$MX$_6$ have been limited to a single surface termination and have employed the HSE functional, which tends to underestimate the band gaps of halide perovskites due to the absence of long-range dielectric screening~\cite{Kaewmeechai_2022, Liu_2024, Garba_2025}.

In this letter, we go beyond these limitations by employing \emph{ab initio} density functional theory (DFT) to characterize multiple surface terminations of Cs$_2$MX$_6$ featuring different metal cations (M = Zr, Sn, Te) and halogens (X = Cl, Br, I), and by mapping their absolute energy levels. We first explore the electronic properties of the bulk materials using a non-empirical dielectric-dependent hybrid functional (DSH0)~\cite{Cui_2018}, which has been shown to accurately predict the band gaps of both 3D and quasi-2D perovskites~\cite{Garba_2025} and, unlike HSE, is designed to capture long-range dielectric screening effects. This bulk-level accuracy enables us to determine in close agreement with available experimental data the absolute ionization potentials (IP) and electron affinities (EA) of the studied surface terminations. Our results indicate that the DSH functional provides accurate prediction of the electronic band gaps across all Cs$_2$MX$_6$ materials, achieving the lowest mean absolute error compared to standard hybrid functionals such as HSE and PBE0. The absolute energy levels of the most stable surface terminations are computed and compared to available experimental results, allowing us to identify potential HTL and ETL candidates based on the intrinsic absolute positions of their electronic energy levels. Taken together, our results provide a comprehensive map of the energy level alignment at VODP surfaces, offering design guidelines for the next generation of VODP-based optoelectronic devices.

\begin{figure}[!ht]
\centering
\includegraphics[width=0.6\columnwidth]{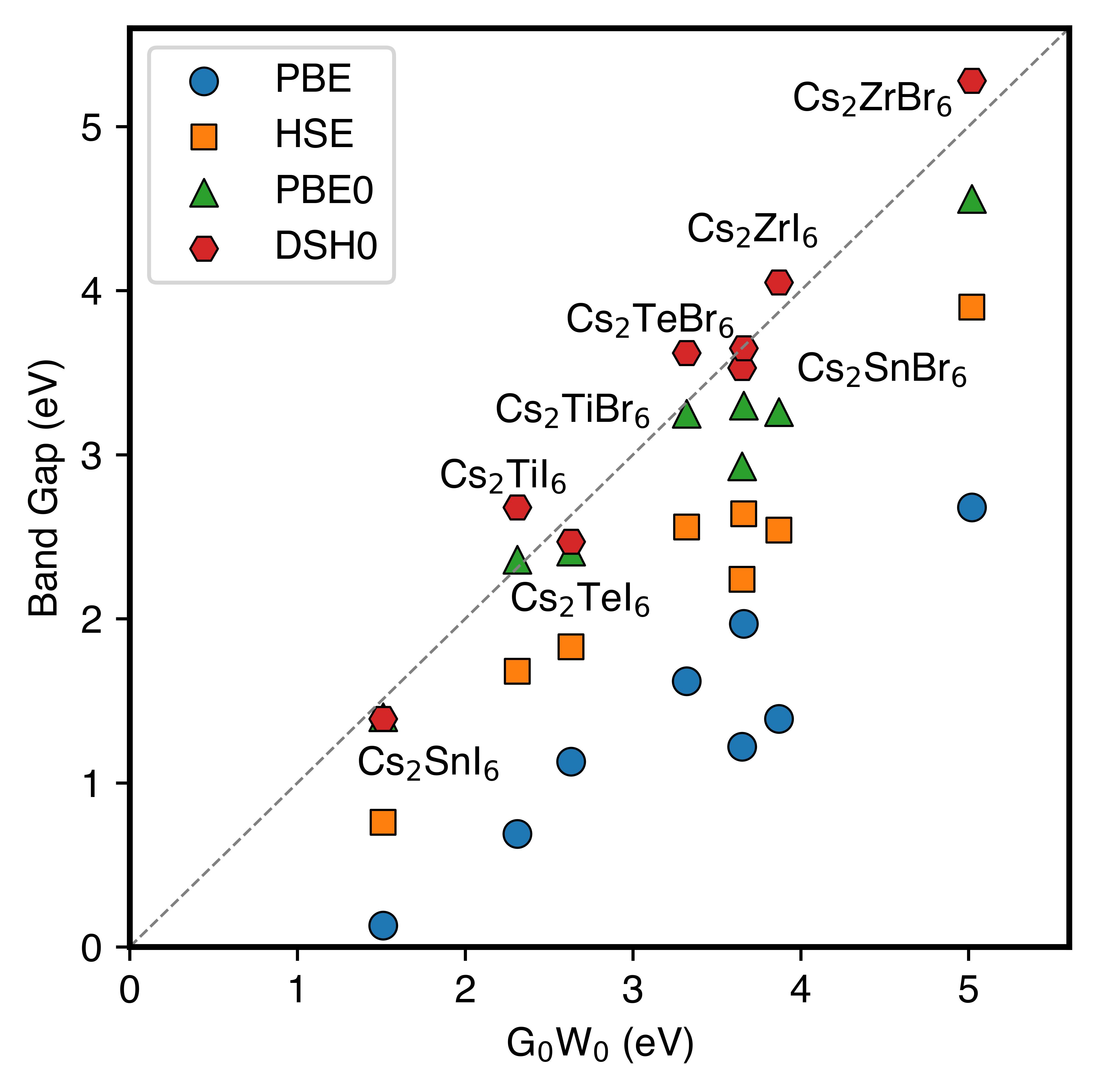}
\caption{\label{fig:Band_gaps} Comparison of the calculated fundamental band gaps of A$_2$MX$_6$ vacancy-ordered double perovskites obtained with  PBE, hybrid HSE and PBE0, and the dielectric-dependent hybrid functional DSH0. Literature G$_0$W$_0$ values from refs.~\citenum{Cucco_2021,Cucco_2023}. }
\end{figure}

We have selected representative Cs$_2$MX$_6$ (X = Cl, Br, I) materials based on the valency of the tetravalent M-site cations, specifically featuring Sn$^{4+}$, Te$^{4+}$, and Zr$^{4+}$, which correspond to $s$, $p$, and $d$ closed shells, respectively. The bulk properties of these materials have been extensively studied ~\cite{Cucco_2021, Cucco_2023, Kavanagh_2022, Ghorui_2024}. Our calculated optimized lattice parameters and bond lengths (Table S1) show excellent agreement with experimental values, differing by less than 2\%. In the following discussion, we focus on representative compounds with M-site cations: Sn$^{4+}$, Te$^{4+}$ and Zr$^{4+}$. Results for other materials (i.e., Ge$^{4+}$, Se$^{4+}$, and Hf$^{4+}$) are included in the Supporting Information (SI). Figure~\ref{fig:Band_gaps} shows the calculated band gaps of VODPs with spin-orbit coupling at various levels of theory, including PBE, HSE, PBE0, and DSH0. While PBE is well known to underestimate band gaps, both HSE and PBE0 also underestimate the band gaps of Cs$_2$MX$_6$ perovskites. DSH0 provides the best agreement with state-of-the-art GW band gaps, with the lowest mean absolute error (MAE) of 0.19 eV, followed by PBE0 with an MAE of 0.32 eV, HSE with an MAE of 0.98 eV and PBE with 1.89 eV. Table S6 shows the values of the computed band gaps. The DSH0 functional, with mixing parameters based on the material dielectric constant, has been shown to successfully predict experimental band gaps of many materials~\cite{Cui_2018}, including halide perovskites~\cite{Garba_2025}. 

VODPs exhibit larger exciton binding energies compared to typical AMX$_3$ perovskites due to the weak interactions between their isolated octahedra~\cite{Cucco_2023, Kavanagh_2022}. Cucco et al.~\cite{Cucco_2023} reported large exciton binding energies in Cs$_2$SnI$_6$ (190 meV), Cs$_2$SnBr$_6$ (510 meV), Cs$_2$ZrI$_6$ (540 meV), Cs$_2$ZrBr$_6$ (780 meV), Cs$_2$TeI$_6$ (530 meV), and Cs$_2$TeBr$_6$ (1240 meV). Therefore, in order to assess the accuracy of the predictions, it is essential to compare the DFT-calculated band gaps with the fundamental electronic band gaps obtained from photoemission experiments or state-of-the-art GW calculations instead of optical measurements~\cite{Maughan_2019}. In this letter, we have chosen to compare our DFT band gaps to the GW band gaps~\cite{Cucco_2021, Cucco_2023}, as illustrated in Figure~\ref{fig:Band_gaps}. Overall, DSH0 provides excellent predictions of band gaps in VODPs. This functional is well suited to materials with homogeneous dielectric screening~\cite{Cui_2018,Chen_2018}, such as conventional 3D halide perovskites~\cite{Garba_2025}, while also naturally capturing the impact of the intrinsic ordered vacancies within the VODP octahedral network on that screening. We therefore move on with DSH0 for the subsequent energy-level calculations in these materials.

\begin{figure}[!htbp]
\centering
\includegraphics[width=1.0\columnwidth]{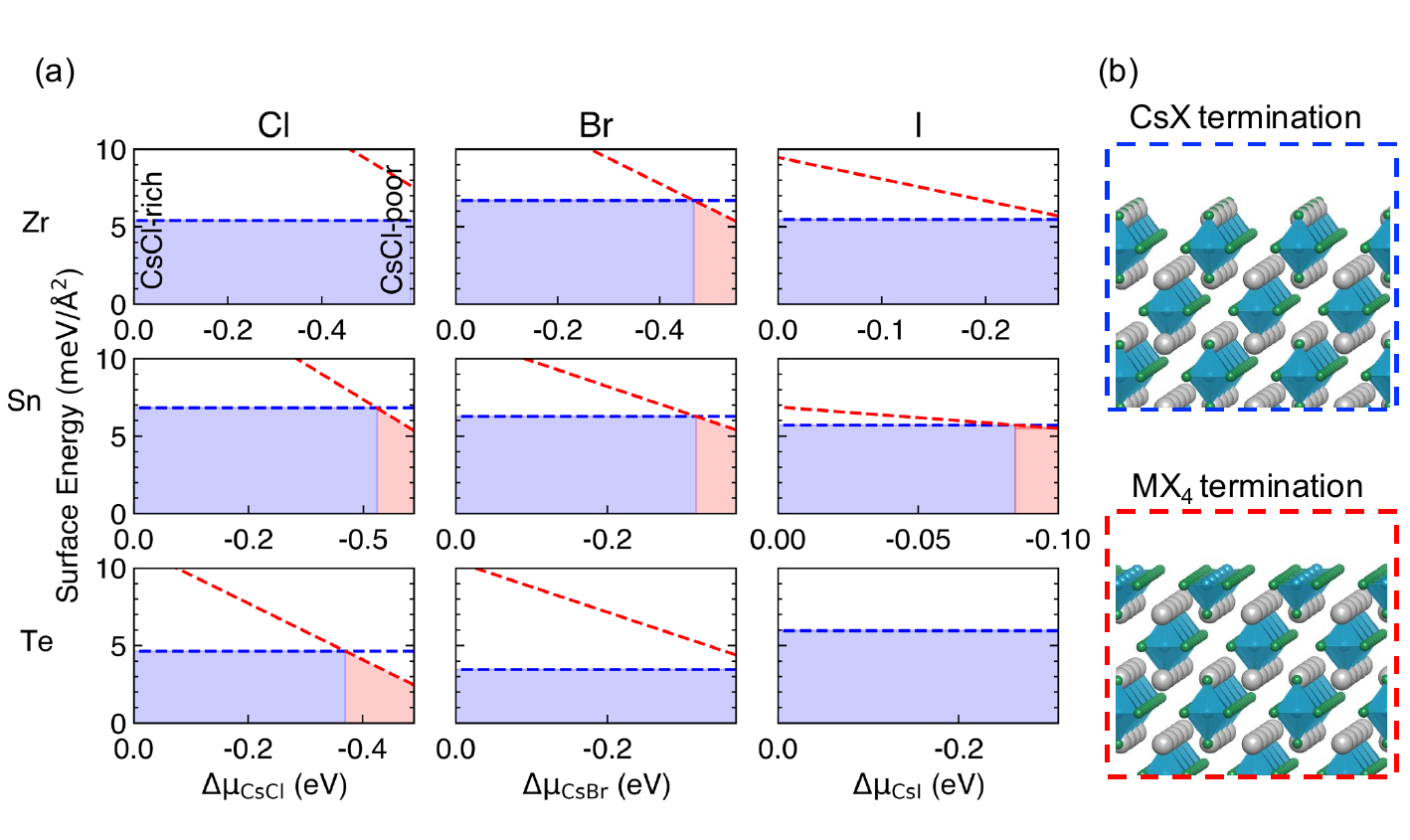}
\caption{\label{fig:fig_2_surfaces_stability} Stability of Cs$_2$MX$_6$(001) surfaces with CsX and MX$_4$ terminations as a function of CsX chemical potential. The Cs$_2$MX$_6$
stoichiometric surfaces show constant energy under both CsX-rich and CsX-poor conditions.}
\end{figure}

We consider surface slab models featuring two different non-polar terminations: CsX and MX$_4$, which correspond to the alternating layers of the Cs$_2$MX$_6$ perovskite structure along the (001) direction, as illustrated in Figure ~\ref{fig:fig_2_surfaces_stability}. We focus on the non-polar (001) facet, consistent with reports identifying it as the most stable surface in related halide perovskites~\cite{Rieger_2023, Li_2023, Jung_Walsh_2017, Haruyama_2014,Volonakis_2018}. Other facets warrant future study, particularly if experimental supporting data emerge. The layer-by-layer composition of these surfaces also matches the composition of the materials synthesis precursors of Cs$_2$MX$_6$, employed in both solution- and vacuum-based synthesis methods~\cite{Maughan_2019, Kumar_2024, Murugan_2025}. For each Cs$_2$MX$_6$ perovskite, we constructed slab models starting from a fully relaxed unit cell with optimized cell volume and atomic positions, featuring seven octahedral layers and a 20~\AA~ vacuum for both CsX and MX$_4$ terminations. This thickness ensures a well converged bulk-like region in the center of the slab and minimizes the interactions across the periodic repetition of the slabs. To evaluate the surface stability we calculate the surface energy with respect to the chemical potential of the respective CsX salt, following the methodology detailed in Ref.~\citenum{Volonakis_2018}.  Figure ~\ref{fig:fig_2_surfaces_stability} shows the surface energy calculated for the two terminations of all studied Cs$_2$MX$_6$ materials. Each row represents the three different M-site cations (i.e., Zr, Sn and Te) within the Cs$_2$MX$_6$, while each column corresponds to Cs$_2$MX$_6$ with the same halogen (i.e., Cl, Br, and I). 

For all Cs$_2$MX$_6$ compounds we find that under CsX-rich conditions  ($\Delta\mu_\text{CsX}=0$), unsurprisingly, the CsX-terminated surfaces have lower surface energy and are therefore predicted to be favorable. On the other hand, under CsX-poor (i.e., MX$_4$-rich) conditions, only some materials have a surface energy of their MX$_4$ termination below that of CsX. In fact, for Cs$_2$ZrCl$_6$, Cs$_2$TeBr$_6$, Cs$_2$ZrI$_6$, and Cs$_2$TeI$_6$, the CsX-terminated surface is favored across the whole range of $\Delta\mu_\text{CsX}$, indicating that for these materials the [001] surface is most likely CsX-terminated regardless of the chemical environment during growth. Moreover, we find that for the remaining materials the crossover from CsX-favored to MX$_4$-favored termination occurs very close to the MX$_4$-rich limit, indicating that the CsX termination is in general more favorable and is most likely the dominant termination present under normal conditions. We note that the surface energies for all terminations in these materials are significantly smaller than those of their three-dimensional double perovskite counterparts~\cite{Volonakis_2018}. We attribute this to the lower energy required to cleave a surface through a network of isolated octahedra in VODPs, compared to that of a corner-sharing octahedral network. We also verify the dependence of the surface energy on different DFT exchange-correlation (xc) functionals in the case of Cs$_2$SnBr$_6$ and observe that, while there is a small variation in the surface energies, all xc functionals show a similar trend, with the CsBr-terminated surface being thermodynamically more stable than the SnBr$_4$-terminated surface (see Figure~S3). 

\begin{figure}[!htbp]
\centering
\includegraphics[width=0.95\columnwidth]{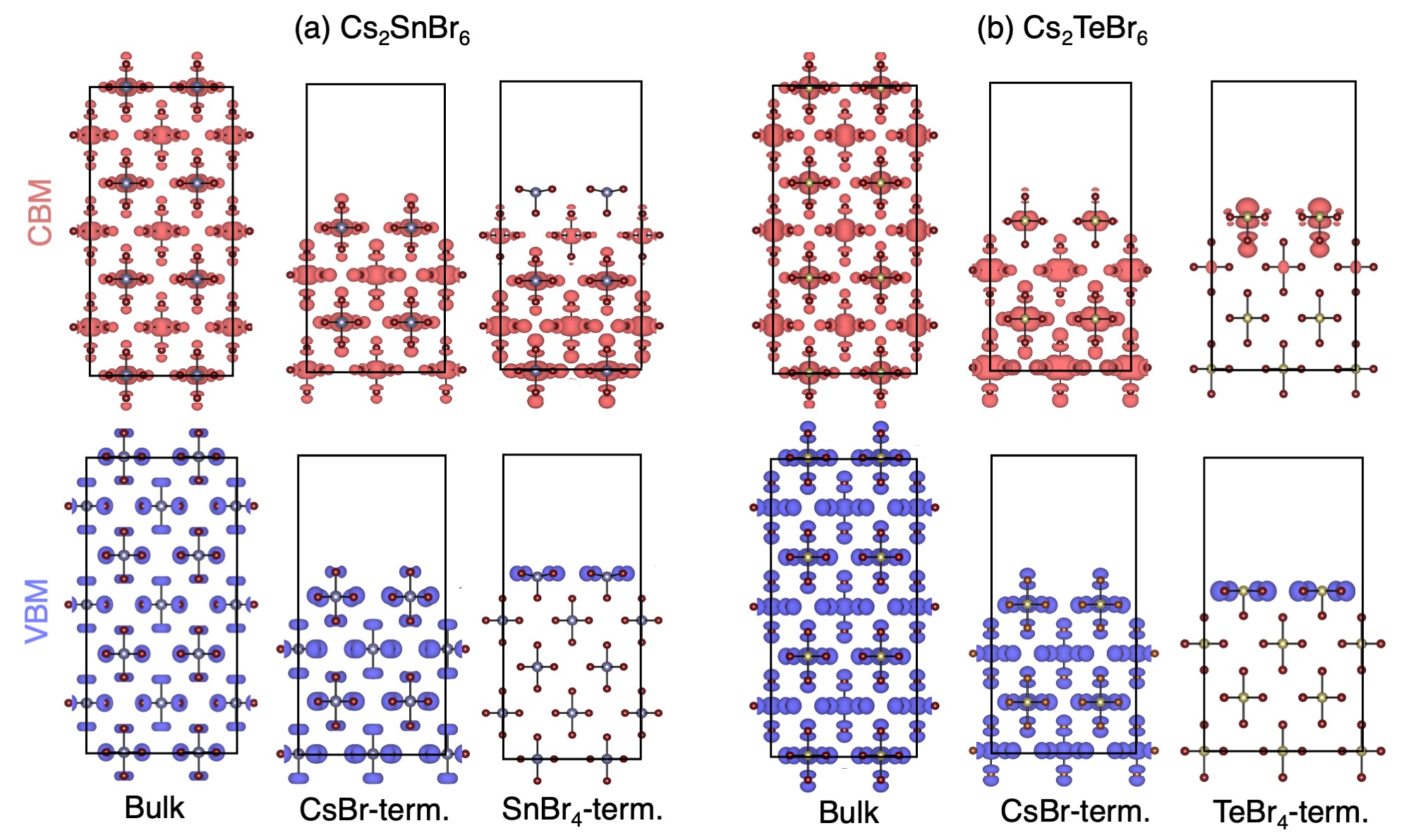}
\caption{\label{fig_3_surface_states} Charge density of the VBM and CBM shown for Cs$_2$MBr$_6$ (M = Sn or Te) in the bulk, the (001) CsBr-terminated surface, and the (001) MBr$_4$-terminated surface. For CsBr-terminated surfaces, the CBM and VBM are quite similar to the bulk, while MBr$_4$-terminated surfaces have surface states localized on Br atoms in Sn-based systems above the VBM, and both VBM and CBM in Te-based systems.}
\end{figure}

The presence of surface states within the band gap is of particular importance, as these states can act as charge carrier traps, promoting non-radiative recombination and thus reduce the performance of these materials within optoelectronic devices. Having established the surface models, we move on to investigate the presence of surface states for the different terminations of Cs$_2$MX$_6$, and calculate the localized pseudo-charge density of the VBM and CBM, as shown in Figure~\ref{fig_3_surface_states} and Figure~S7. A comparison of the bandstructure of the bulk with that of the two terminations is provided in Figure~S4-S6, which also include the relative position of the surface states. The CsX-terminated surfaces exhibit no surface state contributions within the band gap, and the partial charge densities are similar to those of the bulk for all Cs$_2$MX$_6$ compounds (M = Sn, Zr, and Te). In contrast, the MX$_4$-terminated surfaces exhibit an orbital composition of the CBM that is comparable to the bulk; however, at the VBM, a clear surface state composed of halogen $p$ orbitals is observed for Sn- and Zr-based materials. In the case of Te-based materials, the surface state at the CBM is composed of interacting halogen and Te $p$ orbitals. Finally, we employ the DSH0 dielectric-dependent hybrid functional in the case of Cs$_2$SnBr$_6$ slabs to validate the surface states beyond local xc functionals. Our results indicate that surface states persist in both PBE and DSH0 calculations, as shown in Figure~S8 of the SI. Consequently, we report that for CsX terminations there are no surface states, while for MX$_4$ terminations a state related to halogen $p$ orbitals appears above the valence band maximum, while only for Te both band edges are surface related states. Hence, the metal exposure at the MX$_4$ terminations could significantly deteriorate carrier lifetime and mobility compared to the CsX-terminated surfaces.

\begin{figure}[!htbp]
\centering
\includegraphics[width=0.75\columnwidth]{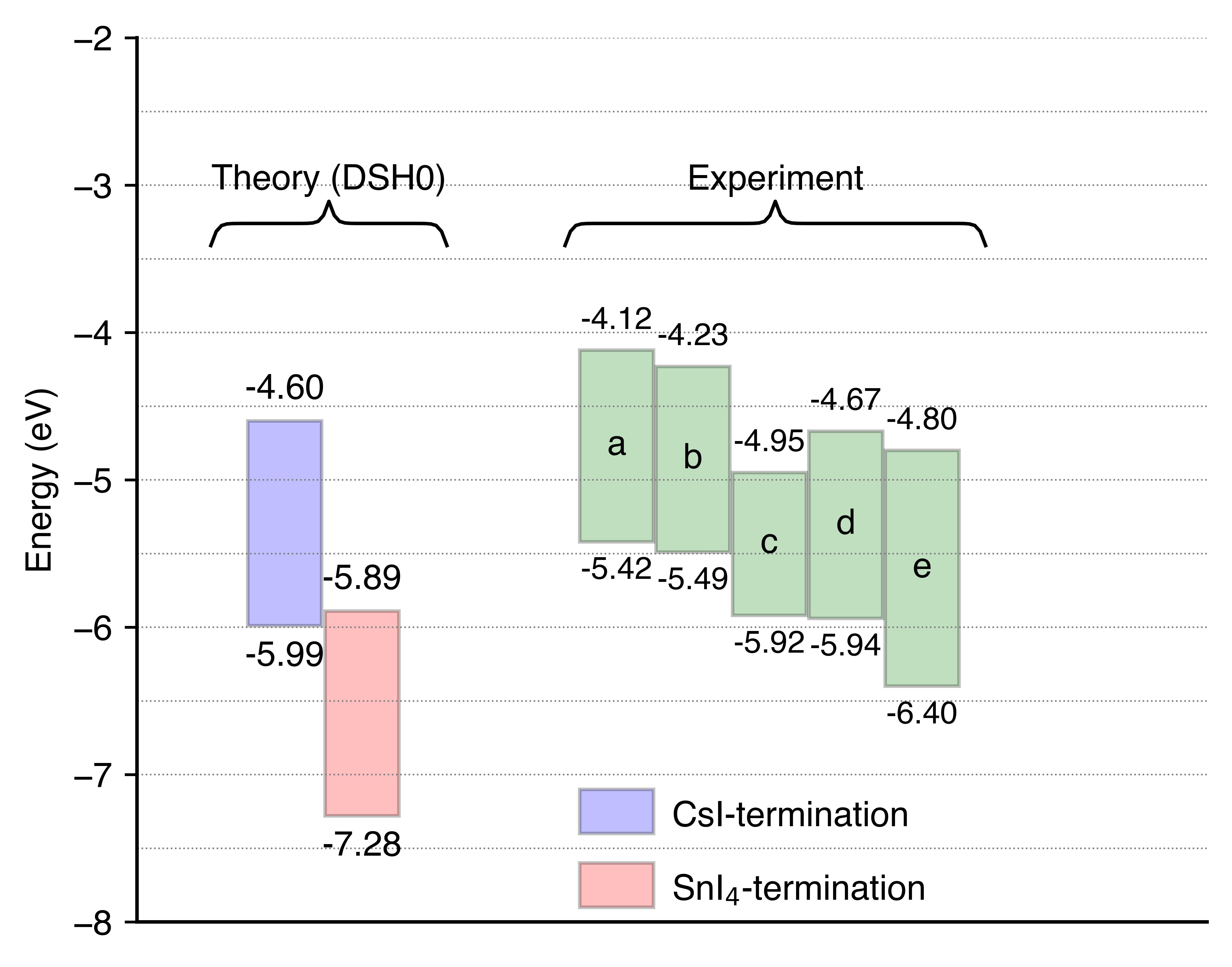}
\caption{\label{fig_3_SnI6} Comparison of absolute energy levels using DSH0 functional and experiment for Cs$_2$SnI$_6$. Experimental values represented by a,b,c,d and e are from Refs.~\citenum{Lee_2017},\citenum{Lee_2014},\citenum{Maughan_2016},\citenum{Zhang_2014}, and \citenum{Saprov_2016}, respectively.}
\end{figure}

\begin{figure}[!htbp]
\centering
\includegraphics[width=1.0\columnwidth]{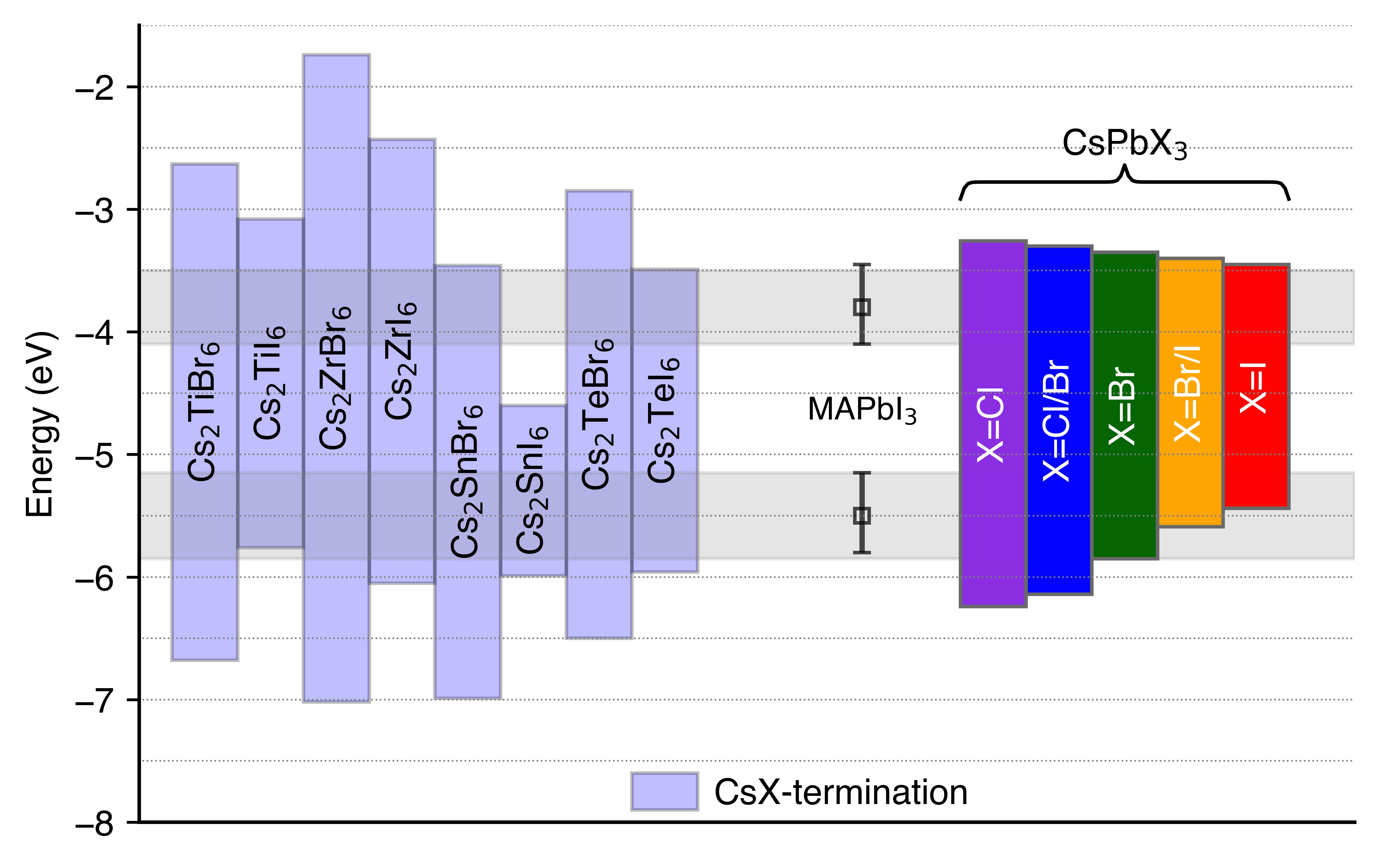}
\caption{\label{fig_4_all_EA_IP} Energy levels of Cs$_2$MX$_6$ (M = Ti, Zr, Sn, Te; X = Br, I) computed using DSH0 functional for CsX-terminated surfaces. Experimental energy levels for MAPbI$_3$ and the error bars are from Ref.~\cite{Schluz_2016, Kim_2015} while that of CsPbX$_3$ nanocrystals are taken from ref.\cite{Ravi_2016} }
\end{figure}

For each Cs$_2$MX$_6$ compound, we calculate the EA and IP by aligning the Kohn-Sham valence band maximum (VBM) and conduction band minimum (CBM) of the bulk systems to the vacuum level of their corresponding surface slabs.~\cite{Van_de_Walle_1987, Hinuma_2014, Volonakis_2018} This procedure (see Figure S9) avoids the computational expense of slab calculations at the hybrid DFT level. The calculated IPs and EAs are given in Table~S7-S8. Having validated the DSH0 approach for calculating band gaps in VODPs, we now assess its performance in predicting the IPs and EAs of these materials. As a representative case, we consider Cs$_2$SnI$_6$, one of the most extensively studied VODPs, and compare DSH0-predicted IP and EA values with available experimental data, as shown in Figure~\ref{fig_3_SnI6}. Overall, the DSH0 values lie within the range of experimentally reported values, validating the performance of the functional in describing the energy level alignment. In particular, the experimental range is best reproduced by the CsI-terminated surface, which is consistent with the favorable surface stability of the CsI termination over the SnI$_4$ termination discussed above (see Figure~\ref{fig:fig_2_surfaces_stability}), thus indicating that the experimental samples are likely to be predominantly CsI-terminated. We also note that the experimental energy levels vary considerably across studies, mainly due to differences in synthesis conditions and measurement techniques~\cite{Kumar_2024, Maughan_2019}. Despite this variability, the DSH0-predicted energy levels show particularly close agreement with the values obtained from UPS/IPES measurements~\cite{Saprov_2016}. Other experimental results, in which the IPs are obtained from photoemission and the EAs are derived by adding the optical~\cite{Lee_2017, Lee_2014, Zhang_2014} or DFT~\cite{Maughan_2016} band gap, also fall within the broader experimental range. These findings further support the reliability of DSH0 in predicting energy levels for Cs$_2$SnI$_6$, and we therefore use this functional to calculate IP and EA values for all other VODPs considered in this work. We also include comparison with experimental data available for Cs$_2$SnBr$_6$ and Cs$_2$TiBr$_6$ in Figure~S10. We also note that comparing with experimental alignments is particularly challenging for materials with large excitonic effects, such as Ti-based VODPs, especially when the measurement involves a shift to the optical band gap.

Figure~\ref{fig_4_all_EA_IP} shows the calculated energy levels of Cs$_2$MX$_6$ (M = Zr, Sn, Te; X = Br, I) for their most stable CsX-terminated surfaces, alongside those of MAPbI$_3$ and prototypical CsPbX$_3$ nanocrystals. We also added the energy levels of Cs$_2$TiBr$_6$ and Cs$_2$TiI$_6$ to illustrate the effect of M-site substitution in materials with the same $d$ closed-shell (see Figure~S11 for MX$_4$ termination). The error bars in the reported experimental values for MAPbI$_3$ indicate the ranges found in the seminal works~\cite{Schluz_2016, Kim_2015}. Among the VODPs studied here, Cs$_2$ZrI$_6$ and Cs$_2$TiI$_6$ exhibit energy level alignments suggesting they could act as HTLs, while Cs$_2$SnBr$_6$ exhibits an optimal EA to function as an ETL for MAPbI$_3$-based devices. Since level alignment alone does not determine suitability as a transport layer, we note that Cs$_2$ZrI$_6$ and Cs$_2$TiI$_6$ also have relatively low hole effective masses (0.6$m_0$), while Cs$_2$SnBr$_6$ has a low electron mass (0.2$m_0$)~\cite{Cucco_2021,Kavanagh_2022,Cucco_2023}, favoring their use as HTL and ETL, respectively. Cs$_2$SnBr$_6$ exhibits optimal energy levels to function as an electron injection layer for CsPbBr$_3$ and CsPb(Br$_x$Cl$_{1-x}$)$_3$ perovskite LEDs. The EA of Cs$_2$SnBr$_6$ is $-3.46$~eV, which is shallower than that of ZnO ($-3.90$~eV) and closely matches that of Mg$_x$Zn$_{1-x}$O ($-3.40$~eV), the latter of which has demonstrated a twofold increase in EQE of CsPbBr$_3$ nanocrystals compared to ZnO~\cite{Wu_2017}. Figure~\ref{fig_4_all_EA_IP} also highlights the potential energy barriers for hole and electron extraction from MAPbI$_3$ and CsPbBr$_3$ when using VODPs as charge transport layers. In more detail, the IPs of the iodides lie higher than those of the bromides across all functionals, with values increasing from PBE to HSE, PBE0, and DSH0. This trend is linked to the widening of the band gap in Cs$_2$MX$_6$ as the halogen ionic radius decreases (i.e., from I to Br to Cl)~\cite{Cucco_2021, Maughan_2019}. For instance, the IPs for the CsX-terminated surface of Cs$_2$SnBr$_6$ and Cs$_2$SnI$_6$ at the DSH0 level are $-6.99$~eV and $-5.99$~eV, respectively, which compare well with the experimental values of $-6.62$~eV and $-5.94$~eV for polycrystalline samples~\cite{Zhang_2014}. A similar, though less pronounced, trend is observed for Cs$_2$TeBr$_6$ and Cs$_2$TeI$_6$, where the IP (EA) values are $-6.50$~eV ($-2.85$~eV) and $-5.96$~eV ($-3.49$~eV), respectively. Furthermore, given the potential for metal miscibility at the M-site and the tunability of VODPs to form high-entropy semiconductors~\cite{Liga_2023,Folgueras_2023, Zhang_2025, Phan_2025}, VODPs offer a promising opportunity as stable inorganic HTL/ETL materials with tunable energy levels, enabling better compatibility with a broader range of absorbers.

In summary, the dielectric-dependent hybrid DSH0 has the best agreement with GW band gaps (MAE of 0.19 eV) and successfully reproduces experimental IPs and EAs for Cs$_2$SnI$_6$, establishing it as the functional of choice for energy level calculations in VODPs. The CsX-terminated (001) surface is thermodynamically more stable than the MX$_4$ termination across most chemical conditions and, crucially, is free of in-gap surface states. In contrast, MX$_4$ terminations introduce halogen $p$-derived surface states above the VBM, and at both band edges for Te-based materials, which can act as non-radiative recombination centers and significantly deteriorate carrier lifetime and device performance. This highlights the special care that should be taken when synthesizing these materials in order to achieve the desired surface termination. Finally, based solely on the intrinsic absolute positions of the computed energy levels, we identify Cs$_2$ZrI$_6$ and Cs$_2$TiI$_6$ as potential HTL candidates and Cs$_2$SnBr$_6$ as a promising ETL for both MAPbI$_3$-based solar cells and CsPbBr$_3$ perovskite LEDs, and we quantify the energy barriers for hole and electron extraction at these interfaces. We note that these conclusions are based purely on energy level alignment and do not account for carrier transport mobility or conductivity, which are additional factors that must be considered in the design of functional devices. The tunability of these energy levels through halide mixing and M-site alloying further broadens the potential applicability of VODPs across a wider range of optoelectronic applications.

\section*{Supplementary Material}
Computational details, Convergence tests for surface slabs and for the vacuum level alignment procedure, Calculation of DSH0 band gaps, Surface stability, Bandstructure of bulk and surfaces, Charge density of the CBM/VBM for bulk and surfaces, Absolute energy levels, Additional comparisons between calculated and experimental band-edge positions for Cs$_2$SnBr$_6$ and Cs$_2$TiI$_6$.

\begin{acknowledgments}
We acknowledge funding from the Agence Nationale pour la Recherche through the CPJ program and the SURFIN project (ANR-23-CE09-0001), the ALSATIAN project (ANR-23-CE50-0030), and the ANR under the France 2030 programme, MINOTAURE project (ANR-22-PETA-0015). We acknowledge computational resources from the HPC resources of TGCC under the Allocation Grant No. 2025 - A0190907682  made by GENCI, and from the EuroHPC Joint Undertaking and supercomputer LUMI, hosted by CSC (Finland) and the LUMI consortium through a EuroHPC Regular Scale Access call.

\end{acknowledgments}
\section*{Author Declarations}
\subsection*{Conflict of interest} The authors have no conflicts to disclose.

\section*{Data Availability Statement}
The data that support the findings of this study are openly available in Materials Cloud Archive [https://doi.org/10.24435/materialscloud:4h-t4].

\bibliography{main}

\end{document}